\documentclass[10pt]{NSP1}
\usepackage{url,floatflt}
\usepackage{helvet,times}
\usepackage{psfig,graphics}
\usepackage{mathptmx,amsmath,amssymb,bm}
\usepackage{float}
\usepackage[bf,hypcap]{caption}
\usepackage{algorithmic}
\usepackage{algorithm}
\usepackage{listings}
\usepackage{multirow}

\emergencystretch=\maxdimen
\def\no{\noindent}

\def\firstpage{67}
\def\thevol{8}
\def\thenumber{2}
\def\theyear{2019}

\begin{document}

\titlefigurecaption{{\large \bf \rm Information Sciences Letters }\\ {\it\small An International Journal}}

\title{Application of Intelligent Multi Agent Based Systems For E-Healthcare Security}

\author{Faizal Khan\hyperlink{author1}{$^1$} and Omar Reyad\hyperlink{author2}{$^{1,2,*}$}}
\institute{$^1$College of Computing and Information Technology, Shaqra University, KSA  \\
           $^2$Computer Science Branch, Faculty of Science, Sohag University, Egypt}

\titlerunning{Multi Agent Based Systems For E-Healthcare Security}
\authorrunning{F. Khan and O. Reyad}

\mail{ormak4@yahoo.com}

\received{25 Feb. 2019}
\revised{23 Apr. 2019}
\accepted{27 Apr. 2019}
\published{1 May 2019}

\abstracttext{In recent years, availability and usage of extensive systems for Electronic Healthcare Record (EHR) is increased. In medical centers such hospitals and other laboratories, more health data sets were formed during the treatment process. In order to enhance the standard of the services provided in healthcare, these records where shared and can be used by various users depends on their requirements. As a result, notable issues in the security and privacy where obtained which should be monitored and removed in order to make the use of EHR more effectively. Various researches have been done in the past literature for improving the standards of the security and privacy in E-health systems. In spite of this, it is not completely enhanced. In this paper, a comprehensive analysis is done by selecting the existing approaches and models which were proposed for the security and privacy of the E-healthcare systems. Also, a novel Intelligent-based Access Control Security Model (IBAC) based on multi agents is proposed to maintain and support the security and privacy of E-healthcare systems. This system uses agents in order to maintain security and privacy while accessing the E-health data between the users.}


\maketitle

\section{Introduction}  \label{intro}
The influence of Information Technology (IT) has induced and combined numerous electronic health information from various places such as laboratories of medical research, hospitals and health care firms \cite{c1,c1a}. This led to the formation of a novel concept called Electronic health (E-health).  This can be defined as the application of IT based technologies and the practices of E-commerce for the entire sharing and processing, of the health information. This information is called as the EHR. The sensors based medical systems can be controlled by various controls in order to record of patient endlessly throughout the medical process. Hence, users like medical staff can access these collected health records since it contains most of the confidential and sensitive health data. Different methodologies have been proposed earlier in for maintaining the privacy and security of the EHR \cite{c2,c2a}. But, these methods need more security in order to distribute the health data. The E-healthcare systems are real-time and has patient information which is in digital format. These are maintained by licenced persons. These data sets where formed by acquiring various data from different patients. In these EHRs, the authorised persons can be the patients or the doctors. The data present in the servers can be available in local or cloud based which stores, analyse the stored health data \cite{c3}. The components which are present in the networks can be the inter connector between the patients and the medical staff for enhancing the broadcasting and distribution of data \cite{c4}. Though there are many benefits in these systems, more treats are there in terms of the security and the privacy for the data used in it. These security threats are caused because of their design [20]. These threats can be classified into various categories such as the data collection level \cite{c5,c6,c7,c8,c9,c10}, transmission level \cite{c11,c12,c13,c14}, and storage level \cite{c15,c16,c17,c18,c19} which are described more clearly in section \ref{Sec2}. Due to these threats in security and privacy of the EHR data, some users are not ready to use these applications. Hence, it is necessary to make sure that the users should be ready to use the system without any hesitation. Therefore, it is important to propose a system for maintaining the privacy and security in the EHR data.

In this paper, a detailed survey is done for analysing the security and privacy threats present in the healthcare system. Then, a novel agent based system for providing the security and privacy for EHR data is also proposed. The agent based system consists of various intelligent agents such as the user interface agent, authentication agent, connection establishment agent and the connection management agent with distinct and expedient working features. These intelligent agents make ease of use, effective communication between patients / users and the E-service providers. Various functions such as interface framing, user registration, user authentication, connection establishment and connection maintenance where carried out by the agent based system. 

The remainder of this paper is organized as follows. Section \ref{Sec1} presents the various types of attacks and the methods proposed earlier to avoid it. Section \ref{Sec2} presents the intelligent agents and the proposed agent based method for providing the privacy and security in the E healthcare systems. Comparison with other methods are discussed in Section \ref{Sec3}. Conclusions and the future work are depicted in Section \ref{Sec4}.

\section{Related Studies} \label{Sec1}
Obtaining security and privacy in E-health data is a crucial part in maintaining the objectives of this advanced methodology. There is a possibility of different forms of attacks while making the health related data to be shared among the users and the medical persons \cite{c11}. Hence, majority of the healthcare institutions have already developed the advanced framework for making the data as highly secured. Various guidelines have been proposed earlier for maintaining the requirements of privacy and security in the field of E-health. Generally, the healthcare systems can be easily accessed by intruders. This sort of unauthorised access affects the overall performance of healthcare systems \cite{c4}. Networks present in the hospitals \cite{c19,c3} can be easily accessed by the hackers. Jamming based attack, collision based attack, de-synchronization based attacks, spoofing and selective forwarding based attacks are the common types of attacks that is done so far during the stage data collection. Threats such as altering the overall information, embedding unwanted data, can be done at the level of data collection. Various security models have been proposed by earlier researchers in order to avoid these types of attacks at the stage of data collection.

Shin et al. \cite{c12} studied the different types of security models proposed already for the health-care platform and tried to find out how to avoid the leakage of information in this health-care platform. The framed various requirements of the security in order to make sure that the overall security and privacy in electronic health are up to the standard. As a solution to the above mentioned problem, the authors used a modified security model called as the Role Based Access Control (RBAC). They proposed a novel architecture by using the improved RBAC model. This model is proposed for maintaining the privacy and security in the health care data mainly at the stage of transferring the data between the user and the health care provider. Their algorithm can be applied in smart devices also. The results obtained revealed that their proposed algorithm provides more security and privacy in securing the health care data.

Morchon and K. Wehrle in \cite{c18} proposed a novel modular based privacy and security mechanism for the healthcare applications. Their algorithm is a modified version of the already available RBAC system. This system is developed because of two main conditions. First one is to assign and share the access control policies of mechanisms proposed for access control to the entire nodes of a sensor. Secondly, for storing the data which are related to the user such as the location, time, information about the user's health. It can be classified into critical condition, emergency condition or normal condition. Various attacks such as altering the patient's medical information, intruding in the middle, data tampering, scrambling, signalling, unfairness in resource allocation, modification of message, data interception can be done in the networks. Other threats such as spying of the data, altering the overall information, interrupting in between the communication, sending additional and unwanted signals to block the source of the message where caused by the attackers at the data transmission level. It is easy for an unauthorised developer to construct the systems for intruding or changing the data of the patient through the wireless based technology. Hence, a mechanism for controlling the overall transfer should be there whenever the patient's information is transferred \cite{c8}.

Fadoua Khennou et al. \cite{c20} proposed a novel framework for secure authentication and transmission of health data using advanced encryption techniques. This methodology is proposed for the security of two purposed such as the data and channel. The security of channel is provided by utilizing the Secure Sockets Layer (SSL) on the Hypertext Transfer Protocol (HTTP) layer present in the networks, while the data security is provided on the Simple Object Access Protocol (SOAP) layer which is present above the HTTP. The authors proposed that their methodology can be used along with multi-factor authentication in order to provide authentication in the health data. Kahani et al. \cite{c3} framed a new scheme of access control in health data. Their methodology is based on a zero-knowledge protocol for verifying and maintaining the the user's identity. Their framework combines the public key and a secret session key which is generated by Derive Unique Key Per-Transaction (DUKPT) scheme in order to establish secure communication between different health entities.

Gajanayake et al. \cite{c17} proposed new mechanism access control for enabling the privacy and security of the E-health data. Their method is a combination of three security based models. Such as the Discretionary Access Control (DAC), Mandatory Access Control (MAC) and the Role Based Access Control (RBAC) \cite{c16} in order to form a novel architecture. Their methodology allows the user as well as the healthcare providers to access the data in a privileged manner. The mail disadvantage of this methodology is it can be used only with in one type of network model while accessing the health data.
Access level based attacks can be a threat for the storage. It can change the health information of a patient, changing the overall configuration of system or servers which is involved in the monitoring purposes. These attacks can be done in the Interference of Patient's Information, accessing of medical information of patient without approval etc. can be done in the storage level of patient's medical records. Various issues based on the hardware and software present in maintaining the patient's medical record can cause an intervention in the healthcare system. Also it is also possible that the health care system can be easily accessed by various types of threats such as viruses, Trojans, and spyware etc. \cite{c19}. Lili Sun and Hua Wang \cite{c2} proposed an improved access control usage and model. Main task of this method is to give the privilege to access private data more securely. It consists of six important components such as, attributes of subject, object attributes, rights, authorizations, obligations, and conditions. In these attributes, the authorizations, obligations, and conditions are components of usage control decisions which are further applied to determine whether a user is allowed to access the data. The presence of other attributes paves the way for solving certain shortcomings that have been common in access controls. The main disadvantage of this model is that it represents only a first step for authorization model.

\subsection{Motivation of The Proposed Framework} \label{Sec1.1}
Challenges faced by the conventional learning techniques are as follows: 

\begin{enumerate}
\item \, Various methods were proposed earlier for improving the EHR. Majority of the EHR systems are still in process, but with poor accuracy.
\item \, Artificial Intelligence (AI) algorithm based methods was utilized for improving the accuracy of the EHR model. Though, it was highly scalable, but the performance was poor in terms of accuracy.
\item \, The methods based on fuzzy systems were proposed later on for the health records. The errors where evaluated. The method was more efficient and robust, but resulted in poor accuracy.
\end{enumerate}

In order to rectify the drawbacks present in the existing methodologies, this framework is proposed.

\section{Intelligent Agents} \label{Sec2}
An agent is an autonomous and flexible computer based system that can receive input and output from the environment. There are many other advantages of the agents. For example, agents can be given knowledge to do specific task, collaborate with each other, and extendible which are very useful when developing secure systems. An agent can represent a user and do tasks on behalf of the user. Moreover, agent systems are extendible in nature. A new agent can be created instantly and added to the existing system in order to represent a new user such as a network monitoring agent, without changing the entire system.

\subsection{Proposed IBAC Security Model} \label{Sec2.1}
\begin{figure*}[h!]
\centering
\includegraphics[width= 0.80\textwidth]{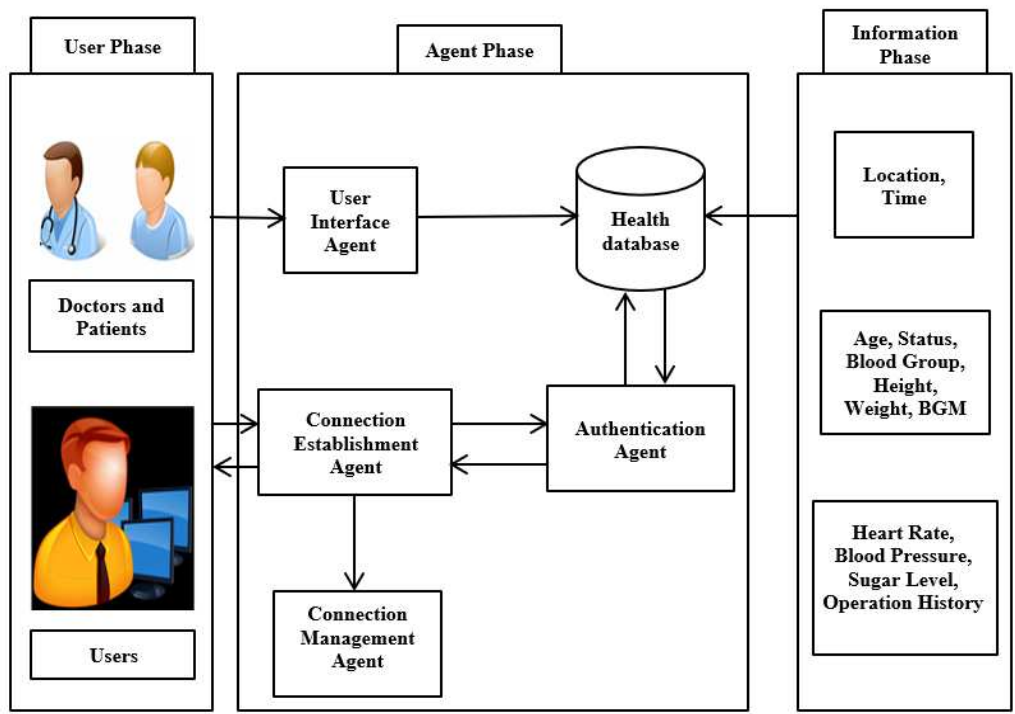}
\caption{Architecture of the Proposed IBAC Model}
\label{fig1}       
\end{figure*}

A Multi-Agent System for providing effective and secure E-health security services has been projected in this section. Architecture of the Intelligent-based Access Control Security Model (IBAC) using the agent based systems in order to enhance the security in the access of healthcare systems is shown in Figure \ref{fig1}. The Multi-agent system proposed in this work consists of various intelligent agents with distinct and expedient working features such as the user interface agent, authentication agent, connection establishment agent and the connection management agent. These intelligent agents make ease of use, effective communication between patients / users and the E-service providers. Various functionalities such as forming the interface, registration of the user, authentication of the user, establishing the connection and maintaining the connection where done by the multi-agent system. It also uses a database in order to store and retrieve the health information in the form of electronic health record.

The proposed methodology consists of three main phases such as the User phase, Agent phase and the Information phase. The user phase consists of all the users who are dealing with this system. Users can be doctors, patients and the authorised professional for handling the health records for further processing. The agent phase consists of multi agents which is responsible for maintaining the security in accessing the health records. The protocols and policies were defined by the user interface agent. A health database is used to store all the username and the passwords provided and used by the patients, physicians and the authorised person for accessing and further processing data. A user authentication agent is used to validate the user. It is also used to verify the user credentials present in the health database. A connection establishment agent along with the connection management agent is used to establish the authenticated connection.

\subsection{Working Principle} \label{Sec2.2}
The proposed system accepts the service request from the customer through the user interface agent. It accepts the username and password of users. The user interface agent is connected through a website or a mobile based application. The authentication agent validates the username and password of the customer. After authenticating the user, a connection is established between the user and the server side by the connection establishment agent. This paves the way for effective and secure access of the E-health care data. The information about the established connection will be also stored in the database so that it can be to use for further references. After establishing the connection, the authentication agent grants the permission for users to access the credentials of the person who requested the data.

The information phase present in the health database consists of the basic and all the necessary information and data regarding the patient. It has the environment information such as location of the patient and time of the data collected. Information regarding the patient such as Age, Status, Blood Group, Height, Weight, BGM etc. where present in the patient information field of the database. Current medical information such as Heart rate, Blood Pressure, Sugar level, Operation history etc. where present in the current medical data field. All the username and the passwords provided and used by the patients, physicians and the authorised person for accessing and further processing data is also stored in the health database.

\begin{lstlisting} [caption = \textbf{Algorithm for connection establishment}, label = list1]
For every user U in the user set UT
    if (UN \in U) and
       AP = <UR, M, FI, FS>
          then,
            Establish connection;
          else
            No connection;
    End if, 
End for
\end{lstlisting}

\begin{lstlisting} [caption = \textbf{Algorithm for accessing the files}, label = list2]
For every UT in U
    Compute the authentication policy
      if (AP = FALSE)
           Break;
      end if;
      if(AP = TRUE), then;
      if FA = <U, M, FI, FS>
          then, 
            Access the files along 
            with all the fields;
      end if;
End for

\end{lstlisting}

\subsection{The Algorithm} \label{Sec2.3}
Health systems have complicated accessing protocols since there exists many actors in the system. The health care system should support many users, their roles, files present in the database and permissions to access the contents present in the health database. In this section, algorithm for authenticating the user, establishing the connection between users as given in \ref{list1} while in \ref{list2} the system, accessing the files for the authorised users algorithm is presented.

\subsection{Research Findings} \label{Sec2.4}
\begin{enumerate}
\item \, The proposed framework based on agent based system can provide an effective and secure E-health security services
 
\item \, This framework is controlled by the patient who is considered a major part of the E-health system

\item \, The two types of intelligent agents such as the user interface agent, authentication agent make ease of use, effective communication between patients / users and the E-service providers

\item \, The user interface agent is connected through a website or a mobile based application so that it will be easy to handle for the users.
\end{enumerate}

\section{Comparison With Existing Methods} \label{Sec3}
In this section, the proposed model for providing security in E-health data is compared with the existing approaches. Table \ref{tab1} shows the performance comparison of various techniques in E-health security.

{\begin{table}[h!]
\caption{Comparison of techniques for various E-health security methods}
\label{tab1}       
\begin{center}
\begin{tabular}{ | p{3.5cm} | p{3.5cm} | } \hline
E-health security methods                   &  Technique used       \\  \hline
CARE Model \cite{c3}	                            &  Modules for each actors and their role  \\ \hline
Patient-Centered Multi Agent System \cite{c8}      &  Agents for individual Users          \\ \hline
Body Sensor Network and Agent based Medical Server \cite{c28}	& Agents for Patient, Supervisor, Doctor and manager  \\  \hline
Biometric based Method \cite{c21}	             & Electrocardiogram and fingerprint for the authentication purpose  \\  \hline
Biometrics based Method \cite{c22}    & Electrocardiogram and photoplethysmography for the authentication purpose.  \\ \hline
Proposed method                   &	Multi Agents for all functionalities            \\  \hline
\end{tabular} 
\end{center}
\end{table}} 

Drawbacks present in the existing methods where rectified in this proposed method using intelligent agent based systems. The performance remains higher in securing the health data. From the Table \ref{tab1}, it can be observed that the proposed method uses multiple agents for all functionalities in the process of securing the E-health data more effectively.

\section{Conclusion} \label{Sec4}
Patient's health information should be kept secure in medical based servers so that the healthcare specialists and the doctors can access and use it at the time of treatments. Attacks at these health data servers can alter the information, modify the overall data and make unauthorised person as an accessed user. In order to make sure the data is secure, a novel agent based technology is proposed in this paper. The proposed security model IBAC aims to avoid these using the intelligent agent based systems. This model is novel since it uses the intelligent agents for the entire process of providing access control and sharing. The proposed model consists of agents, each of which is in charge for different task. This method is a simple and efficient access control mechanism based on the agents functionalities. Future work can be the implementation of the proposed model with more agents in real-time health datasets and also to calculate its efficiency and accuracy.


\emergencystretch=\hsize

\begin{center}
\rule{6 cm}{0.02 cm}
\end{center}


\begin{floatingfigure}[h]{3.5cm}
\centering
\includegraphics[width=2.5cm,height=2.5cm]{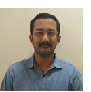}
\end{floatingfigure}
{\bf \hypertarget{author1}{Faizal Khan}} is currently working as an assistant professor at the Department of Computer Science in the College of Computing and Information Technology (CCIT), Shaqra University, the Kingdom of Saudi Arabia. He has published more than 50 articles in referred journals and acted as an editorial member and reviewer in many reputed journals and conferences. His research interests include Image processing, Pattern Recognition, Intelligent based Systems and Medical image analysis.

~

\begin{floatingfigure}[h]{3.5cm}
\centering
\includegraphics[width=2.5cm,height=2.5cm]{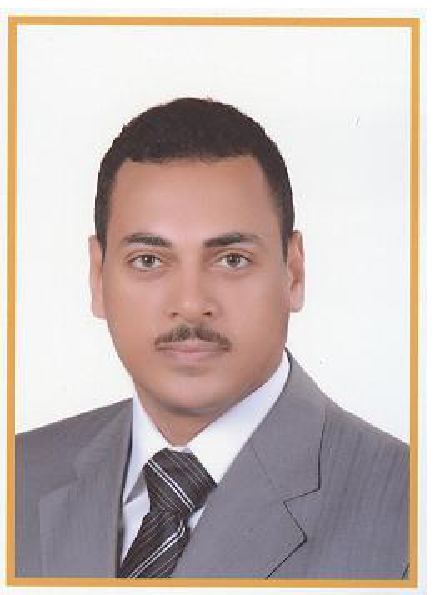}
\end{floatingfigure}
{\bf \hypertarget{author1}{Omar Reyad}} is currently a Lecturer of Computer Science at Sohag University, Egypt. He received his PhD in Informatics from the Faculty of Electronics and Information Technology, Warsaw University of Technology, Poland. He received his MSc in Computer Science from Sohag University, Egypt. He is the author and co-author of over 25 research papers on Elliptic curve cryptography, Cryptographic protocols, Biometric security and Chaos-based cryptography.


\end{document}